\documentclass{article}

\def\DH{\rm I\kern-1.5pt\rm H\kern-1.5pt\rm I}

\input epsf

\def\DR{\rm I\kern-1.45pt\rm R}
\def\DC{\kern2pt {\hbox{\sqi I}}\kern-4.2pt\rm C}

\newcommand{\nn}{\nonumber}
\newcommand{\ba}{\begin{array}}
\newcommand{\ea}{\end{array}}
\newcommand{\be}{\begin{equation}}
\newcommand{\ee}{\end{equation}}
\newcommand{\bea}{\begin{eqnarray}}
\newcommand{\eea}{\end{eqnarray}}
\newcommand{\bi}{\begin{itemize}}
\newcommand{\ei}{\end{itemize}}

\newcommand{\p}[1]{(\ref{#1})}

\begin{document}\thispagestyle{empty}
\vspace{2cm}
\begin{flushright}
\end{flushright}\vspace{2cm}
\begin{center}
{\Large\bf Dual multiplets in N=4 superconformal mechanics}
\end{center}
\vspace{1cm}

\begin{center}
{\large\bf Stefano Bellucci${}^{a}$, Sergey Krivonos${}^{b}$ and Anton Sutulin${}^{b}$ }
\end{center}

\begin{center}
${}^a$ {\it
INFN-Laboratori Nazionali di Frascati,
Via E. Fermi 40, 00044 Frascati, Italy} \vspace{0.2cm}

${}^b$ {\it
Bogoliubov  Laboratory of Theoretical Physics, JINR,
141980 Dubna, Russia} \vspace{0.2cm}

\end{center}
\vspace{2cm}

\begin{abstract}
\noindent We propose Lagrangian formulations of a
three-particles translation invariant $N=4$ superconformal mechanics
based on the standard and twisted $(1,4,3)$ supermultiplets. We
show that in the appropriate set of coordinates, in which the
bosonic kinetic terms for each of the two cases take
conformally-flat forms, the corresponding models produce just
similar physical systems with the flipped angular part of bosonic potential
$U(x)\rightarrow 1/U(x)$. This flipping looks so simple only being
written in the ``angle'' variable, while in the standard variables
it looks more complicated to preserve the superconformal symmetry.
We demonstrate at both the superfield and the component level, how
familiar $N=4$ supersymmetric 3-particles models, including
$3-Calogero$, ${\it BC_2}$ and ${\it G_2}$ ones, can be constructed with twisted
supermultiplets. We also present some new explicit examples of
3-dimensional superconformal mechanics.
\end{abstract}

\newpage
\setcounter{page}{1}

\setcounter{equation}0
\section{Introduction}

The recent interest in the supersymmetric extension of conformal mechanics \cite{AFF} is motivated by
its connection with the four-dimensional theory including $N=2$ SUGRA in the context of black hole
physics \cite{CLAUS,GIBBON,Mohaupt}, as well as with the superextension of some one-dimensional
integrable systems \cite{CAL,OP1,OP2,WCM,Haschke}.

In the case of one-dimensional $N=4$ supersymmetry there is a rich
set of supermultiplets each containing four fermionic degrees of
freedom but a different number of physical bosons (for a review
see e.g. \cite{SBSK}). They admit $N=4$ superconformal invariance
under the general $D(2,1;\alpha)$ superconformal group
\cite{Sorba} in the standard superspace, as it was shown in many
papers \cite{donets,IKL2,IKL1}.

A particular class of $N=4$ superconformal mechanics, which arises
in superextensions of integrable models such as the Calogero one,
has been intensively investigated recently
\cite{Gal,BGK,BKS-CAL,FIL,FIL1,FIL2}. Studying these models has
led to a connection between them and a system of non-linear
differential WDVV equations \cite{BGL,GLP,GLP1,GLP2,KLP,LP}.

Most of these superconformal models were based on the standard $N=4$ supermultiplet \cite{leva}
whose physical boson was recognized as a dilaton field.
This supermultiplet is transformed as a vector, under the action of the $SU(1,1|2)$ superconformal group,
which is in agreement with the rule of conservation of defining constraints under superconformal transformations.
On the other hand, there exits another twisted (or "mirror") supermultiplet \cite{Pashnev} with the same component field
content as the standard one, but with different properties under superconformal transformations of the same group.
This supermultiplet was also described in \cite{delduc2} in the framework of Harmonic Superspace (HSS) \cite{HSS},
which was formulated for the one-dimensional case in \cite{IL,delduc1}.

In our previous consideration of the three-particles $N=4$
supersymmetric $Calogero$ model \cite{BKS-CAL} we used the standard
$N=4$ supermultiplet and constructed within it the superconformal
extension of some integrable models in superspace, for a
particular choice of the parameter $\alpha=-1$ that corresponds to
the $SU(1,1|2)$ superconformal group. The main subject of the
present paper is to demonstrate the equivalence of three-particles
models with translational isometry (or two-particles ones without
the latter) whose superfield actions are based either on standard
\cite{leva} or twisted \cite{Pashnev} $(1,4,3)$ supermultiplets.
The existence of such an equivalence is not surprising, because
the translation invariance, together with the (super)conformal
one, fixed the 3-particle action, up to one arbitrary function
which defined the model completely. Then one may properly choose
this arbitrary function to reproduce some desired system. The most
interesting feature of the approach with different supermultiplets
is that starting with the (almost) same superspace action and
performing just the replacement {\it standard} $\leftrightarrow$
{\it twisted} superfield one can obtain similar systems with
flipped angular part of bosonic potential  $U(x)\leftrightarrow 1/U(x)$. Thus, these
interchanging of the supermultiplets is indeed related with the
duality transformation.

We start with the description of the realization of the
$SU(1,1|2)$ superconformal group in $N=4$ superspace \cite{IKL1}
and find how the standard and twisted supermultiplets are
transformed under the action of this group demanding the
preservation of the defining constraints. In Section 3 we
construct $SU(1,1|2)$ superconformal invariant actions for both of
these supermultiplets in superspace. Then, passing to the on-shell
form of each of the actions, we demonstrate that, in a proper set
of new bosonic and fermionic variables, the corresponding actions
lead to similar three-particles models with the flipped angular part of bosonic
potential $U(x)\rightarrow 1/U(x)$. We also show how to choose the
metric functions (super potentials), as well as the coupling
constants, to reproduce the same systems  within the approach with
standard and twisted supermultiplets. Finally, we give some
explicit examples of (super)conformal invariant  models, including
the $3-Calogero$, ${\it BC_2}$ and ${\it G_2}$ ones, as well as some new models.

\setcounter{equation}0
\section{$SU(1,1|2)$ superconformal invariant action}
In the present paper we consider the $N=4$ superconformal
mechanics based on two supermultiplets $V$ and $\Phi$, which are
known as standard \cite{leva} and twisted \cite{Pashnev} ones,
respectively. Each of them has the same component content
$(1,4,3)$ and their general supersymmetric actions lead to similar
models at the component level. We will call these multiplets dual,
because the resulting on-shell actions are related via a duality
transformation of the bosonic potential $U\leftrightarrow 1/U$. On
the other hand, the superconformal properties of the multiplets
are different. In this section we show that $V$ transforms as a
vector under the $SU(1,1|2)$ supergroup, while $\Phi$ is a scalar
under the action of the latter.

\subsection{$SU(1,1|2)$ superconformal group in superspace and super-dilaton}
We consider the real $N=4$ superspace
$\mathbf{R}^{(1|4)} = (t, \theta_a, \bar \theta^a)$,
in which the spinor derivatives satisfy the standard anticommutation relations
\be\label{alg}
\left\{ D^a, \bar D_b \right\}=2i\delta^a_b
\partial_t, \quad \left\{ D^a, D^b \right\}=\left\{ \bar D_a, \bar
D_b \right\}=0.
\ee
In this superspace one may  realize the most general  $D(2,1;\alpha)$ $N=4$ superconformal group \cite{Sorba}.
In what follows, we will deal only with the special case of $D(2,1;\alpha)$ with $\alpha=-1$, which corresponds
to the $SU(1,1|2)$ superconformal group. This supergroup acts on the coordinates of $\mathbf{R}^{(1|4)}$ as \cite{IKL1,leva}
\be\label{conf-tran}
\delta t = E -\frac{1}{2} \theta_a D^a E -\frac{1}{2}\bar\theta^a \bar D_a E,
\qquad
 \delta \theta_a=-\frac{i}{2} \bar D_a E,\; \delta
\bar\theta^a=-\frac{i}{2} D^a E,
\ee
where the introduced superfunction
$E(t,\theta,\bar\theta)$ collects all parameters of $SU(1,1|2)$ transformations
\be\label{E}
E=f(t)-2i \left( \varepsilon_a \bar\theta^a
-\theta_a \bar\varepsilon^a \right) + \theta^a \bar\theta{}^b B_{(ab)}
+ 2\left( \dot\varepsilon_a \bar\theta^a
+\theta_a \dot{\bar\varepsilon}^a\right)(\theta\bar\theta) +\frac{1}{2}
(\theta\bar\theta)^2 \ddot{f}.
\ee
Here, the bosonic $f(t)$ and fermionic $\varepsilon^a(t)$
functions are restricted to be
\be\label{param}
f=a+bt+ct^2,\quad \varepsilon^a = \epsilon^a + t \zeta^a.
\ee
In \p{E}, \p{param} the bosonic parameters $a,b,c$ and $B_{(ab)}$ correspond to
translations, dilatations, conformal boosts and rigid $SU(2)$
rotations, while the fermionic ones $\epsilon^a$ and $\zeta^a$ are the parameters
of Poincar\`e and conformal supersymmetries, respectively.

It is rather important that the spinor derivatives transform through themselves under
the $SU(1,1|2)$ defined in \p{conf-tran}, as \cite{IKL1,leva}
\be\label{tran-der}
\delta D^a =\frac{i}{2} \left( D^a \bar D_b E\right)D^b, \quad
\delta \bar D_a = \frac{i}{2}\left( \bar D_a D^b E\right) \bar D_b\;.
\ee

In order to construct a superconformal invariant action, one has to take into account
that the superspace integration measure, defined as
\be\label{mera}
ds = dt d^2 \theta d^2 \bar\theta \;,
\ee
transforms, in accordance with \p{conf-tran}, under the $SU(1,1|2)$ as
\be\label{mera-tran}
\delta ds = - \partial_t E \; ds\;.
\ee
Therefore, in order to maintain $SU(1,1|2)$ invariance, one has to introduce
the super-dilaton~--~$N=4$,~$d=1$ superfield~$Y$
which transforms as follows\footnote{Sometimes, the name super-dilaton is used for the superfield $X=Log Y$.}
\be\label{propY}
\delta Y = \partial_t E \cdot Y.
\ee
Now, one defines the superconformal invariant measure in the following way
\be\label{mera-inv}
\triangle s = Y ds.
\ee
The super-dilaton $Y$ has to be further constrained by the irreducibility conditions
\be\label{dilaton2}
\left[D^a, \bar D_a\right] Y=0.
\ee
These constraints are fully compatible with the transformation properties \p{tran-der}, \p{propY}.

As follows from \p{dilaton2}, the super-dilaton $Y$ contains the
set of component fields, that is $(1,4,3)$, defined by
\be\label{Y}
y =Y|, \quad \lambda^a = i D^a Y, \quad  \bar
\lambda_a = i \bar D_a Y, \quad Y_{(ab)} = D_{(a} \bar D_{b)} Y.
\ee
Besides them, there is also an additional constant parameter
$M$ arising as a corollary of the constraints \p{dilaton2}
\be\label{M} \frac{\partial}{\partial t} D^2 Y = 0, \quad
\frac{\partial}{\partial t} \bar D^2 Y = 0 \qquad \Rightarrow
\qquad D^2 Y = i M, \quad \bar D^2 Y = - i M.
\ee
Since this real constant is not of our interest at present, we will take it equal
to zero.

The invariant action for the super--dilaton $Y$ is unambiguously
restored to be
\be\label{action-dilaton} S_{\,\rm Y} = -\int dt
d^2 \theta d^2 \bar\theta \; Y \log{Y}.
\ee
Having the invariant
measure \p{mera-inv}, the superconformal invariant action for
every matter superfield, say ${\cal A}$, can be constructed as
\cite{BKS-CAL}
\be\label{scalarA} S=-\int \triangle s\; G({\cal A}),
\ee
provided the superfield ${\cal
A}$ is a scalar under superconformal transformations.

To describe a three particle system with translation invariance one has to
introduce three superfields: $Y_0$ for the center of mass
coordinate, the super-dilaton $Y$, and one matter superfield
${\cal A}$. Clearly, the translation invariance results in the
decoupling of $Y_0$ from the remaining coordinates. Therefore, to
provide the super-conformal invariance the superfield $Y_0$ has to
be a super-dilaton one obeying \p{propY}, \p{dilaton2}. The second
super-dilaton field $Y$ is needed to construct a superconformal
invariant action for the matter field \p{scalarA}. Thus, the most
freedom we have is the choice of the matter superfield $\cal A$
with the only property to be a scalar under \p{conf-tran}.
Summarizing, we conclude that the most general three-particles
action possessing translation and superconformal invariance reads
\be\label{genaction}
S= -\int dt d^2 \theta d^2 \bar\theta
\;\bigl( Y_0 \log{Y_0}+Y \log{Y} +  Y\; G({\cal A}) \bigr).
\ee
In what follows, we will deal mostly with the two last terms in
\p{genaction} due to the obvious triviality of the action for the
center of mass.

\subsection {The standard and twisted $(1,4,3)$ supermultiplets of $N=4$ mechanics}
As explained above, the direct way of constructing the invariant
superconformal action is to use, together with the super-dilaton
$Y$, some superfield ${\cal A}$ which is a scalar under the
$SU(1,1|2)$ supergroup. The $N=4$ standard \cite{leva} and twisted
\cite{Pashnev} supermultiplets with the component contents
$(1,4,3)$, denoted, respectively, as $V$ and $\Phi$ can be
described by the following constrained superfields in $N=4$
superspace $\mathbf{R}^{(1|4)}$
\footnote {The description of these supermultiplets in the framework of the harmonic superspace
approach was given in \cite{delduc2,delduc1}.}
\be\label{conV}
[D^a, \bar D_a] V =0,
\ee
\be\label{conP} D^{(a} \bar D^{b)} \Phi = 0.
\ee
One can check that the constraints \p{conV} and \p{conP}
are invariant under the $N=4$ superconformal group $SU(1,1|2)$ if
the superfields $V$ and $\Phi$ transform as:
\be\label{prop}
\delta V = \partial_t E \cdot V\,, \qquad \delta \Phi = 0.
\ee
{}From \p{prop} one concludes that the superfield $V$, similarly to the
super-dilaton $Y$ one, has a vector-type transformation under the
action of the $SU(1,1|2)$ supergroup, while $\Phi$ has a scalar
one. Thus, one may directly use the superfield $\Phi$ as the
matter superfield with the action \p{scalarA}, while in the case
of $V$ one has  introduce the following scalar quantity including
the super-dilaton $Y$ \cite{BKS-CAL}
\be\label{Z}
Z=\frac{V}{Y}\quad \Rightarrow\quad S=-\int dt d^2 \theta d^2
\bar\theta \; Y\; G(Z).
\ee

As follows from \p{conV} and \p{conP}, each of these superfields
contains one physical boson, four fermions and three bosonic auxiliary components.
A convenient definition of the components can be taken as
\bea\label{all-comp}
&&
z = Z|, \quad
\phi = \Phi|\,, \\
&&
\psi^a = i D^a Z, \quad \bar \psi_a = i\bar D_a Z, \quad
\xi^a = i D^a \Phi, \quad \bar \xi_a = i \bar D_a \Phi, \nn\\
&& A_{(ab)} = D_{(a} \bar D_{b)} V, \quad B = D^a D_a \Phi, \quad
\bar B = \bar D_a \bar D^a \Phi, \quad C = -\frac{i}{4}\;[D^a,
\bar D_a] \Phi. \nn
\eea
There is an additional constant
parameter, which may be present, in virtue of constraints
\p{conV}, in $V$
\be D^2 V = i m_v, \quad \bar D^2
V = -i m_v.
\ee
In what follows, this parameter should be
identified with the coupling constant of the models considered
below. Note that, in contrast to the superfield $V$, the multiplet
$\Phi$ does not contain a similar constant quantity. In that case
it arises after adding to the general action the Fayet-Iliopoulos
terms (FI-terms) (next Section), or performing the dualization
procedure to one of the auxiliary components entering $\Phi$.

\setcounter{equation}0
\section{On the equivalence of the component actions}
In this section we show that the invariant $SU(1,1|2)$ superconformal models based on
standard and twisted supermultiplets describe the same physical systems.
After obtaining the corresponding component actions, one finds how one of them goes into another,
by the proper change of variables and vice versa.

\noindent
Let us start with the superconformal action of the standard multiplet which is represented
by the superfield $Z$ \p{Z} \cite{BKS-CAL}
\be\label{SUSY-Z}
S_{(\rm Y,Z)} = - \frac{1}{2}\,\int ds\, \Big \{ Y \log Y +
Y F(Z) \Big \}.
\ee
After integrating over $\theta$'s and using \p{all-comp}, one gets the off-shell component action
\footnote {The superspace measure is defined to be: $d^2 \theta d^2 \bar \theta = \frac{1}{16}\, D^a D_a \bar D_a \bar D^a$}
\bea\label{off1}
S_{(\rm Y,Z)} &=& \frac{1}{32}\, \int dt \Big \{ \frac{4}{y} \dot y^2
+ \frac{4i}{y} (\dot \lambda^a \bar \lambda_a - \lambda^a \dot {\bar\lambda}_a)
- \frac{2}{y} Y^2 - \frac{4}{y^2} \lambda_a Y^{(ab)} \bar \lambda_b
- \frac{2}{y^3}\lambda^2 \bar \lambda^2 \nn\\
&+& 4y F'' \dot z^2 + 4iy F''(\dot \psi^a \bar \psi_a - \psi^a \dot {\bar\psi}_a)
- 4i F'' \dot z (\psi^a \bar \lambda_a - \lambda^a \bar \psi_a) \nn\\
&-& \frac{1}{y} F'' m^2_v + \frac{2i}{y} F''
(\psi^a \lambda_a - \bar \lambda_a \bar \psi^a)m_v
- i F'''(\psi^2 - \bar \psi^2)m_v \nn\\
&-& \frac{2}{y} F'' (A^{(ab)} - zY^{(ab)}) (A_{(ab)} - zY_{(ab)})
-4F'' \psi_a Y^{(ab)} \bar \psi_b \nn\\
&-& \frac{4}{y} F''
\Big [\psi_a (A^{(ab)} - zY^{(ab)}) \bar \lambda_b
+ \lambda_a (A^{(ab)} - zY^{(ab)}) \bar \psi_b\Big ]
+ 4F''' \psi_a (A^{(ab)} - zY^{(ab)}) \bar \psi_b \nn\\
&-& \frac{2}{y} F'' (\lambda^2 \bar \psi^2 + \psi^2 \bar \lambda^2)
- \frac{8}{y} F''\lambda^a \psi_a \bar \lambda_b \bar \psi^b
+ 4 F''' (\lambda^a \psi_a \bar \psi^2 + \psi^2 \bar \lambda_a \bar \psi^a)
- y F^{(4)} \psi^2 \bar \psi^2
\Big \}\,, \nn
\eea
where
\be
\left. F'' = \frac{\partial^2 F}{\partial Z \partial Z} \right|_{\theta=0},
\ee
and the following rules of summation are used: $\lambda^2 = \lambda^a \lambda_a,
\bar \lambda^2 = \bar \lambda_a \bar \lambda^a$, etc.
Eliminating the auxiliary bosonic fields $A_{(ab)}, Y_{(ab)}$ and making the following change of variables
\bea\label{newset1}
&&
F'' = 4 \left (\frac{dx}{d z} \right )^2 = h^2(x), \quad u = \frac{1}{2}\, \log y, \nn\\
&&
\lambda^a = 2e^u \eta^a, \quad \bar \lambda_a = 2e^u \bar \eta_a, \quad
\psi^a = \frac{2e^{-u}}{h} \gamma^a, \quad \bar \psi_a = \frac{2e^{-u}}{h} \bar \gamma_a,
\eea
one obtains the on-shell Lagrangian in which the kinetic bosonic part is of the conformally-flat form,
while the kinetic terms for fermions has the standard structure of a free action
\bea\label{Lagr1}
{\cal L}_{stand} &=&  \frac{1}{2}e^{2u} (\dot u^2 + \dot x^2)
+ \frac{i}{2} (\dot \eta^a \bar \eta_a - \eta^a \dot{\bar \eta}_a)
+ \frac{i}{2} (\dot \gamma^a \bar \gamma_a - \gamma^a \dot{\bar \gamma}_a)
-i \dot x (\gamma^a \bar \eta_a - \eta^a \bar \gamma_a) \\
&-& \frac{1}{32}\, e^{-2u}\, h^2 m^2_v + \frac{i}{4}\, e^{-2u}\, h  \Big
(\gamma^a \eta_a - \bar \gamma_a \bar \eta^a \Big) m_v\,
- \frac{i}{8}\, e^{-2u}\, h_x  \Big (\gamma^a \gamma_a - \bar \gamma_a \bar \gamma^a \Big) m_v \nn\\
&-& \frac{e^{-2u}}{4}\, \Big (\gamma^2 \bar \eta^2 + \eta^2 \bar \gamma^2 \Big )
+ \frac{e^{-2u}}{2}\,\frac{h_x}{h} \Big (\gamma^a \eta_a \bar \gamma^2 +\gamma^2 \bar \gamma_a \bar \eta^a \Big )
- e^{-2u} \eta^a \gamma_a \bar \eta_b  \bar \gamma^b \nn\\
&+& \frac{e^{-2u}}{4}\, \bigg [3 + \frac{h^2_x}{h^2}
- \frac{h_{xx}}{h} \bigg ]
\gamma^2 \bar \gamma^2 - \frac{e^{-2u}}{4} \eta^2 \bar\eta^2\,.  \nn
\eea
The superconformal action of the twisted multiplet $\Phi$ can be written as the following integral in $N=4$ superspace
\be\label{SUSY-P}
S_{(\rm Y,\Phi)} = \frac{1}{2} \int ds \Big \{ - Y \log Y +
Y G(\Phi) \Big \}\,.
\ee
It produces, after integrating over Grassmann variables with the help of \p{all-comp}, the off-shell component action
\bea\label{off2}
S_{(\rm Y,\Phi)} &=& \frac{1}{32} \int dt \Big \{ \frac{4}{y} \dot y^2
+ \frac{4i}{y} (\dot \lambda^a \bar \lambda_a - \lambda^a \dot {\bar\lambda}_a)
- \frac{2}{y} Y^2 - \frac{4}{y^2} \lambda_a Y^{(ab)} \bar \lambda_b
- \frac{2}{y^3}\lambda^2 \bar \lambda^2 \\
&+& 4y G'' \dot \phi^2 + 4iy G'' (\dot \xi^a \bar \xi_a - \xi^a \dot {\bar\xi}_a)
- 4i G'' \dot \phi (\xi^a \bar \lambda_a - \lambda^a \bar \xi_a) \nn\\
&+& y G'' (B \bar B - 4C^2)
- 2 G'' (\xi^a \lambda_a \bar B + \bar \lambda_a \bar \xi^a B)
- 4i G''(\xi^a \bar \lambda_a + \lambda^a \bar \xi_a) C \nn\\
&-& 4G'' \xi_a Y^{(ab)} \bar \xi_b
- y G''' \xi^2 \bar B - y G''' \bar \xi^2 B - 4iy G''' \xi^a \bar \xi_a C \nn\\
&+& 2 G''' \xi^2 \bar \xi_a \bar \lambda^a + 2 G''' \xi^a
\lambda_a \bar \xi^2 + y G^{(4)} \xi^2 \bar \xi^2 \Big \}\,,
\eea
where
\be \left. G'' = \frac{\partial^2 G}{\partial \Phi \partial
\Phi} \right|_{\theta=0}\,.
\ee
In contrast with the previous case
we have no coupling constant in the action \p{off2}, because the
invariance of the basic constraints for the superfield $\Phi$
\p{conP} under superconformal symmetry forbids the appearance of
any constant among its components. Therefore, the unique way to
have the coupling constant in the action is to introduce an
additional FI-term. Thus, in the case of the twisted
supermultiplet our basic action \p{genaction} has to be modified
as
\be\label{modaction}
S \rightarrow S+ S^{FI}_{(Y,\Phi)}.
\ee
In general, one should add a different kind of FI-terms to the action
$S_{(\rm Y,\Phi)}$, each of them associated with all possible
auxiliary fields present in it. But in order to preserve the
superconformal invariance of the modified action \p{modaction} one
has to choose the FI-terms in the form in which only the special
combination of complex conjugated auxiliary fields is present
(here $m_{\phi}$ is an arbitrary real constant):
\be\label{FI}
S_{(\rm Y,\Phi)}^{\rm FI} = \frac{i}{32}\, \int dt\, m_{\phi}
\left (B -\bar B\right ).
\ee
Indeed, one may easily check that
the integrand in \p{FI} transforms as a full time derivative under
the $SU(1,1|2)$ superconformal group. Now, excluding the auxiliary
fields, in accordance with their equations of motions, in the
action \p{modaction} and passing to following set of variables
\bea\label{newset2}
&&
G'' = 4 \left (\frac{dx}{d \phi} \right )^2 = g^2(x), \quad u =\frac{1}{2} \log y, \nn\\
&& \lambda^a = 2e^u \eta^a, \quad \bar \lambda_a = 2e^u \bar
\eta_a, \quad \xi^a = \frac{2e^{-u}}{g} \rho^a, \quad \bar \xi_a =
\frac{2e^{-u}}{g} \bar \rho_a.
\eea
one finally obtains the
on-shell Lagrangian with the coupling constant $m_{\phi}$:
\bea\label{Lagr2} {\cal L}_{tw} &=&  \frac{1}{2}e^{2u} (\dot u^2 +
\dot x^2) + \frac{i}{2} (\dot \eta^a \bar \eta_a - \eta^a
\dot{\bar \eta}_a) + \frac{i}{2} (\dot \rho^a \bar \rho_a - \rho^a
\dot{\bar \rho}_a)
-i \dot x (\rho^a \bar \eta_a - \eta^a \bar \rho_a) \\
&-& e^{-2u}\, \frac{m_{\phi}^2}{32g^2} + ie^{-2u}\, \frac{m_{\phi}}{4g} \Big
(\rho^a \eta_a - \bar \rho_a \bar \eta^a \Big)\,
+ i e^{-2u} \frac{m_{\phi}}{8} \frac{g_x}{g^2} \Big (\rho^a \rho_a - \bar \rho_a \bar \rho^a \Big) \nn\\
&-& \frac{e^{-2u}}{4}\, \Big (\rho^2 \bar \eta^2 + \eta^2 \bar \rho^2 \Big )
- \frac{e^{-2u}}{2}\,\frac{g_x}{g} \Big (\rho^a \eta_a \bar \rho^2 +\rho^2 \bar \rho_a \bar \eta^a \Big )
- e^{-2u} \eta^a \rho_a \bar \eta_b  \bar \rho^b \nn\\
&+& \frac{e^{-2u}}{4}\, \bigg [3 -\frac{g^2_x}{g^2} +
\frac{g_{xx}}{g} \bigg ] \rho^2 \bar \rho^2 - \frac{e^{-2u}}{4}
\eta^2 \bar\eta^2\,.  \nn
\eea
The analysis of the explicit
structure of the Lagrangians given in \p{Lagr1} and \p{Lagr2}
leads to the statement that the $SU(1,1|2)$ superconformal models
for standard and twisted multiplets are equivalent if the
following relations between metric functions and coupling
constants take place
\be\label{equivav}
h(x) = \frac{1}{g(x)}\,,
\qquad m_{\phi} = m_v.
\ee
Also, the fermionic component should be identified: $\gamma = \rho$.

Finally, let us note that one can also proceed to the Hamiltonian
analysis of these systems constructing the corresponding momenta,
supercharges, etc. The result of doing so, shows us that the
Hamiltonians coincide, when the rules \p{equivav} are taken into
account.

\setcounter{equation}0
\section{Examples of Superpotentials}
\subsection{Some examples of 3-particles models}
As an equivalence of the superconformal models was obtained for
the actions of standard and twisted $(1,4,3)$ multiplets, we have
a reason to reproduce some known examples of three-particles
systems such as the $Calogero$ \cite{CAL}, ${\it BC_2}$ (with $g_1=g_2
\equiv g$) \cite{OP1,OP2} and ${\it G_2}$ (with $f=g \equiv
g$) \cite{WCM} ones.

Let us remind \cite{BKS-CAL} that the bosonic part of the actions
for the following models - $Calogero$ \cite{CAL}, ${\it BC_2}$ (with
$g_1=g_2 \equiv g_{\rm BC}$) \cite{OP1,OP2} and ${\it G_2}$ (with $f=g
\equiv g_{\rm G}$) \cite{WCM} models in the set of coordinates
$(u,x)$ can be written as
\bea
S_{\rm Cal} &=&\int dt \left[
\frac{1}{2} \sum_{i=1}^3{\dot
x}{}_i^2 -\sum_{i<j} \frac{2g}{(x_i-x_j)^2}\right]= \int dt\, \left[ \frac{1}{2} {\dot X}{}_0^2
+ \frac{1}{2} e^{2u} (\dot u^2 + \dot x^2) -  e^{-2u} \frac{9g}{\cos^2 3x} \right],\label{A1}\\
S_{\rm G_2} &=& \int dt \left[ \frac{1}{2} \sum_{i=1}^3{\dot
x}{}_i^2 -\sum_{i<j} \frac{2g}{(x_i-x_j)^2}-\sum_{i<j; i,j\neq k} \frac{6g}{(x_i-x_j+2 x_k)^2}\right]=\nn\\
&&\int dt\, \left [ \frac{1}{2} {\dot X}{}_0^2+\frac{1}{2} e^{2u} (\dot u^2 + \dot x^2)
-  e^{-2u} \frac{36g}{\sin^2 6x} \right ],\label{A3}\\
S_{\rm BC_2} &=&\int dt \left[ \frac{1}{2} \sum_{i=1}^2{\dot
y}{}_i^2 -g\left( \frac{1}{(y_1 -y_2)^2}+\frac{1}{(y_1 +y_2)^2}+\frac{1}{2 y_1^2}+\frac{1}{2 y_2^2}\right)\right] =\nn\\
&&\int dt\, \left [\frac{1}{2} e^{2u} (\dot u^2 + \dot x^2) -  e^{-2u} \frac{8g}{\sin^2 4x} \right ].\label{A2}
\eea

Here, in the cases \p{A1} and \p{A3}
\be\label{EUCL}
X_0=\frac{1}{\sqrt{3}}\left(x_1+x_2+x_2\right), \quad
y_1=\frac{1}{\sqrt{6}}\left(2x_1-x_2-x_3\right), \quad
y_2=\frac{1}{\sqrt{2}}\left(x_2-x_3\right),
\ee
and, finally,
\be\label{polar}
y_1=e^u \sin{x},\quad y_2=e^u \cos{x}.
\ee

As shown in \cite{BKS-CAL}, the supersymmetric version of these
actions (with a decoupled center of mass in the cases of \p{A2}
and \p{A3}) can be constructed with the pair of superfields
$(Y,Z)$ - the standard $(1,4,3)$ multiplets, when the
corresponding superfunctions $F(Z)$ are taken as
\bea
F_{\rm Cal} &=& \frac{2}{9} \Big [ (1+ Z) \log(1+ Z) + (1- Z) \log(1- Z)\Big ], \quad g=\frac{m_v^2}{648} \nn \\
F_{\rm G_2} &=& \frac{1}{18} \Big [ (1+ 4Z) \log(1+ 4Z) + (1- 4Z) \log(1- 4Z)\Big ],\quad g=\frac{m_v^2}{648} \nn\\
F_{\rm BC_2} &=& \frac{1}{8} \Big [ (1+ 4Z) \log(1+ 4Z) + (1- 4Z)
\log(1- 4Z)\Big ],\quad g=\frac{m_v^2}{64}.
\eea
Since the
corresponding Lagrangians depend on the superfield $Z$, which
transforms as a scalar under the $SU(1,1|2)$ superconformal group,
we can replace this superfield with $\Phi$ and change the overall
sign, in agreement with \p{SUSY-Z}, \p{SUSY-P}. Thus, with these
particular choices of superpotentials $G(\Phi)$ in \p{SUSY-P} we
obtain the following result for the bosonic part of the actions,
after passing to polar coordinates $(u,x)$ \p{newset2}
\footnote{Here $\phi$ denotes the physical bosonic component of
the superfield $\Phi$.}
\bea
S_{\rm Cal}& =& \int dt\, \Big \{\frac{1}{2}\, e^{2u} (\dot u^2 + \dot x^2)
- 9 g\; e^{-2u}\; \cos^2 3x \Big \}, \quad \phi = \sin 3x,\label{P1}\\
S_{\rm G_2} &=& \int dt\, \Big \{\frac{1}{2}\, e^{2u} (\dot u^2 + \dot x^2)
- 36 g \; e^{-2u}\;  \sin^2 6x \Big \}, \quad \phi = \frac{1}{4} \cos 6x, \label{P3}\\
S_{\rm BC_2}& =& \int dt\, \Big \{\frac{1}{2}\, e^{2u} (\dot u^2 +
\dot x^2) - 8 g\; e^{-2u}\;   \sin^2 4x \Big \},\quad \phi =
\frac{1}{4} \cos 4x. \label{P2}
\eea
In order to reconstruct the
actions in the set of Euclidean coordinates $(x_i, i =1,2,3)$, one
has to perform the same change of variables as in \p{EUCL},
\p{polar}.

It is also easy to find the structure of the superpotentials
$G(\Phi)$ which correspond to the original $\it{Calogero}$, $\it
{G_2}$ and  $\it{BC_2}$models:
\bea
G_{\rm Cal}(\Phi) &=& \frac{4}{9}\, \log \left ( 1+ e^{-\frac{9}{2}\Phi}\right ), \quad e^{\frac{9}{2}\phi}
= \tan^2\left(\frac{3}{2}x +\frac{\pi}{4}\right), \label{F1}\\
G_{\rm G_2}(\Phi) &=& \frac{1}{9}\, \log \left ( 1+
e^{-\frac{9}{2}\Phi}\right ), \quad e^{\frac{9}{2}\phi} =\tan^2(3 x), \label{F3}\\
G_{\rm BC_2}(\Phi) &=& \frac{1}{4}\, \log \left ( 1+ e^{-2\Phi}\right ), \quad e^{2\phi} = \tan^2 (2x).  \label{F2}
\eea

Summarizing, we conclude that the superpotential having the same
dependence on standard or twisted supermultiplets leads to the
conformal three-particles models, the potential terms of which are
inverse to each other. However, one should stress that the inverse
form of the potential appears only in the ``angle'' variable $x$.
Indeed, for example for the $3-Calogero$ model we will have
\be
U_{Y,Z}=\frac{9}{2}e^{-2u} \frac{1}{\cos^2(3x)} \quad \rightarrow
\quad U_{Y,\Phi}=\frac{9}{2}e^{-2u} \cos^2(3x).
\ee
The same ``flipping'' of the potential, being rewritten in the standard
$x_i$ coordinates \p{EUCL}, \p{polar} will read
\be\label{flip}
U_{Y,Z}=\frac{1}{(x_1-x_2)^2}+\frac{1}{(x_1-x_3)^2}+\frac{1}{(x_2-x_3)^2}\quad
\rightarrow \quad
U_{Y,\Phi}= \frac{729}{16}\frac{(x_1-x_2)^2(x_2-x_3)^2(x_3-x_2)^2}{(x_1^2+x_2^2+x_3^2-x_1
x_2 -x_1 x_3 -x_2 x_3)^4}.
\ee
Thus, systems which look quite
different are related by a special duality transformations. So,
one may expect that integrability on the one side (say, for the
$3-Calogero$ model) will lead to the integrability on other side,
where the potential looks completely different.

\subsection{Generating function of three-particles models}

The construction considered above has been connected with a
certain kind of superpotential $F(Z)$ or $G(\Phi)$. Let us analyze
which three-partial models can be obtained, proceeding from a
generating function of the following kind
\be\label{GF1} F(Z) =
\alpha \Big [ (1+ \beta Z) \log(1+ \beta Z) + (1- \beta Z) \log(1-
\beta Z) \Big ]
\ee
for the models based on the standard
multiplets.
This superfunction contains parameters
$\alpha$ and $\beta$  which can take any arbitrary values. So,
for example, considering the case of \p{GF1}, we find for the
second derivatives of $F(Z)$ taken at the point $\theta = 0$ the
following expression
\be
F''(z) = 2\alpha \beta^2
\frac{1}{1-(\beta z)^2}\;.
\ee
Performing the following change of variables
\footnote{On equal footing, one can choose $\beta z =
\sin(nx)$.}
\be\label{CH1}
\beta z = \cos(nx), \qquad \rightarrow
\qquad \frac{dz}{dx} = - \frac{n}{\beta}\, \sin(nx), \qquad n\in
\mathbf{N},
\ee
one finds for the bosonic part of the kinetic
terms
\bea\label{LK1}
L^{\rm kin} &=& \frac{1}{32}\, \Big (
\frac{4}{y} \dot y^2 + 4y F'' \dot z^2 \Big ) = \frac{1}{2} e^{2u}
\dot u^2 + \frac{1}{8} e^{2u}\left.  \left [ \frac{2\alpha
\beta^2}{1-(\beta z)^2} \right ] \right |_{\beta z = \cos(nx)}
\left (\frac{dz}{dx} \right)^2 {\dot x}^2 \nn\\
&=& \frac{1}{2} e^{2u} \left (\dot u^2 + \left [ \frac{\alpha
n^2}{2} \right ] {\dot x}^2 \right ),
\eea
where, as well as
above, $u = \frac{1}{2}\, \log y$. Thus, the bosonic kinetic term does not
depend on $\beta$ and takes the conformally-flat form, if the
following relation is satisfied
\be\label{AA1}
\alpha n^2 = 2.
\ee
On the other hand, the potential term depends on the parameter
$\beta$, and has the following expression in new coordinates
\be\label{LP1}
L^{\rm Pot} = - \frac{e^{-2u}}{32}\, m^2_v \left.
F''(z)\right |_{\beta z = \sin(nx)} = - \frac{e^{-2u}}{32}\,
\frac{2\alpha \beta^2}{\cos^2(nx)}\, m^2_v.
\ee
{}From the last
expression one concludes that the value of $\beta$ determines only
the relation between the parameter $m^2_v$ and the coupling
constant $g$ of the corresponding model.
Thus, in principle, one may put it equal to {\bf one}: $\beta=1$.

Summarizing, we will get that the values of the parameters $\alpha, n$ and the corresponding three-particles models
with translational isometry (or the two-particles ones without the latter) are interconnected \vspace{0.5cm}\\

\begin{tabular}{|c|c|c|c|c|c|c|c|}
n&2&3&4&5&6&$\cdots$&$\forall n\in \mathbf{N}$ \\
$\alpha$&1/2&2/9&1/8&2/25&1/18&$\cdots$&$2/n^2$ \\
${\it models}$&-&${\it Cal}$&${\it BC_2}$&-&${\it G_2}$&$\cdots$&- \\\vspace{0.5cm}
\end{tabular}

if $\sqrt{\frac{2}{\alpha}} = n, n\in \mathbf{N}$.

These examples lead us to conclude that, in order to achieve at
the component level the action of three-particles conformal
invariant models, whose kinetic part is just a free action and the
potential one is an arbitrary function of one of the coordinates,
one can start with the superfield Lagrangian, as given in \p{GF1},
without any restrictions on the parameter $\alpha$.
Then, requiring that this parameter satisfies some additional
conditions like those in \p{AA1}, one obtains the
component actions which describe some $N=4$ superconformal
translational invariant three-particles models (or two-particles
ones without such an invariance). In particular, all previously
discussed models, including ${\it A_2 Calogero}$, ${\it BC_2}$ and
${\it G_2}$ \cite{BKS-CAL}, are contained in the superpotentials \p{GF1}
by the appropriate value of the parameter $\alpha$.

\section{Conclusion}
In the present paper we analyzed $N=4$ supersymmetric mechanical
models based on the standard and twisted multiplets, each of which
has $(1,4,3)$ component field content. In both cases we
constructed the superfield actions invariant under a particular
choice of the $D(2,1;\alpha)$ superconformal group with
$\alpha=-1$, which corresponds to the $SU(1,1|2)$ one. We
demonstrated that, at the component level, these actions
correspond to the same three-particles superconformal models, when
the proper relation between the metric functions is performed and
the coupling constants are proportional to each other. We also
obtained the explicit form of the superpotential of some
translational invariant three-particles models (or two-particles
one without such an invariance), including, in particular, the
${\it A_2 Calogero}$ one, as a proper function of the twisted
supermultiplet. Moreover, a new structure of superpotentials, both
for the standard and twisted $(1,4,3)$ supermultiplets, which
leads to a class of three-particles models, was found. In general,
these superpotentials depend on one real parameter $\alpha$, and
after fixing its value one gets the correct expression for the
superconformal extension of the corresponding models.

One of the most interesting features of the constructed models is
the ``flipping'' of the angular part of bosonic potential. If we start from the
system with the standard superfields $Y,Z$ and then replace the
second scalar superfield $Z$ by a twisted one $\Phi$, then the
resulting system will have just a ``flipped'' angular part of bosonic potential
$U(x)\rightarrow 1/U(x)$. But this flipping looks so simple only
being written in the ``angle'' variable, while in the standard
variables it looks more complicated \p{flip}. It is worth to
stress that the main property of the started system -- $N=4$
superconformal symmetry -- is preserved by this duality
transformation. Just this superconformal symmetry on both sides
makes it impossible to flip the potential in the standard
coordinates $x_i$, because $U(x_i)$ and $1/U(x_i)$ have different
dilaton weights. That is why the correcting factor (the power of
the denominator) appears in the explicit form \p{flip}. It is an
interesting question to check whether the dual potential in
\p{flip} is still integrable, as it happens for the $3-Calogero$
model.

The present consideration is just the first step in a more
ambitious task, i.e. to consider the system with $n$ multiplets of
one type and $m$  twisted ones. We hope that the additional
freedom, which is related with the presence of two types of
supermultiplets in the same action, will help to override the old
problem of the construction of four- (and higher) particles
Calogero models with $N=4$ superconformal symmetry
\cite{GLP}-\cite{LP}.

\section*{Acknowledgements}
A.S. is grateful to the Laboratori Nazionali di Frascati for hospitality.
This work was partially supported by the grants RFBF-09-02-01209 and 09-02-91349,by Volkswagen
Foundation grant I/84 496 as well as by the ERC Advanced Grant no. 226455, \textit{``Supersymmetry, Quantum Gravity and Gauge Fields''%
} (\textit{SUPERFIELDS}).

\end{document}